\documentclass[aps,prl,preprint,groupedaddress]{revtex4}
\usepackage{amsfonts}
\usepackage{amssymb}
\usepackage{graphicx}
\newcommand{\ds}{\displaystyle}
\newcommand{\qbq}{\mbox{$q\bar{q}$}}
\newcommand{\qvec}{\mbox{\boldmath $q$}}
\newcommand{\pvec}{\mbox{\boldmath $p$}}
\newcommand{\rvec}{\mbox{\boldmath $r$}}
\newcommand{\jvec}{\mbox{\boldmath $j$}}
\newcommand{\etal}{\mbox{\it et al.}}
\def\bm#1{\mbox{\boldmath$#1$} }

\begin{document}
\title{Magnetization screening from gluonic currents and
scaling law violation in the ratio of magnetic
form factors for neutron and proton}
\author{Murat M. Kaskulov}
 \email{kaskulov@pit.physik.uni-tuebingen.de}
\affiliation{Physikalisches Institut, Universit\"at  T\"ubingen,
		 D-72076 T\"ubingen, Germany}
\author{Peter Grabmayr}
 \email{grabmayr@uni-tuebingen.de}
\affiliation{Physikalisches Institut, Universit\"at  T\"ubingen,
		 D-72076 T\"ubingen, Germany}
\date{\today}

\begin{abstract}
  The ratio~$\mu_pG_E^p/G_M^p$ exhibits a decrease for 
  four-momentum transfer~$Q^2$ increasing beyond 1~GeV$^2$ indicating
  different spatial distributions for charge and for magnetization inside the
  proton. One-gluon exchange currents can explain this behaviour.  The $SU(6)$
  breaking induced by gluonic currents predicts furthermore that the ratio of
  neutron to proton magnetic form factors ${\mu_p}G_M^n/{\mu_n}G_M^p$ falls
  with increasing $Q^2$.  We find that the experimental data are consistent
  with our expectations of an almost linear decrease of the ratio
  $\mu_pG_M^n/\mu_nG_M^p$ with increasing $Q^2$, supporting the statement that
  the spatial distributions of magnetization are different for protons and for
  neutrons.
\end{abstract}

\maketitle

The electromagnetic (e.m.) structure of the nucleon is currently subject to a
  renewed theoretical research.  At four-momentum transfer $Q^2>1$~GeV$^2$,
  this interest is motivated by the recent measurements at JLab of the ratio
  $R_p=\mu_pG_E^p/G_M^p$ between the electric $G^p_E(Q^2)$ and magnetic
  $G^p_M(Q^2)$ form factors of the proton. While the previous extractions of
  the two form factors relied on the Rosenbluth separation, and thus on the
  scaling laws:
\begin{equation}\label{scaling} 
	G_E^p(Q^2) = G_M^p(Q^2)/{\mu_p} = G_M^n(Q^2)/{\mu_n} = G_{D}(Q^2) 
\end{equation} 
  using the dipole form factor
\begin{equation}\label{DPL}
G_{D}(Q^2)= [1 + Q^2/0.71\mbox{GeV}^2]^{-2}
\end{equation}
the recent analysis is based on the measured recoil proton polarization in
   elastic scattering of polarized electrons up to
   $Q^2\sim$5.5~GeV$^2$~\cite{protonratio}.  Methodologically the discrepancy
   between the results from Rosenbluth separation and polarization
   measurements is a quite interesting problem. A global re-analysis of the
   elastic electron-proton scattering data by Ref.~\cite{Arrington:2003df}
   confirms a self-consistent interpretation of the individual cross section
   data when using the Rosenbluth technique, but the inconsistency with
   polarization data is not resolved. An attractive idea which can potentially
   solve this problem was recently proposed~\cite{Guichon:2003qm}.  The
   remarkable decrease of the ratio~$R_p$ from unity indicates not only a
   significant deviation from the simple scaling law but also from the
   non-relativistic constituent quark model (NRCQM). The new data show that
   the charge and magnetization inside the proton is distributed differently,
   which could arise~\cite{KG} through one-gluon exchange (OGE) currents
   dominant at high $Q^2$.

By now several calculations for the proton ratio~$R_p$ within different
  hadronic models are
  available~\cite{Lu1,Miller,GlozmanRiska,Cardarelli,GlozmanBoffi,MillerFrank}.
  We refer to the recent review~\cite{Gao:2003ag}, where one can find a
  discussion of the most recent calculations which agree reasonably well with
  the trend of the experimental data and which will allow for predictions at
  higher $Q^2$ than presently accessible.  We only note, that the
  implementation of relativity is an common feature of all these models and
  all emphasize the necessity of relativistic effects generated by both
  kinematical and dynamical $SU(6)$ breaking for the interpretation of the
  decrease of the ratio~$R_p$.

In this paper we wish to discuss other ratios of form factors in order to find
 further support for the scaling law violation. In particular, we will
 investigate the ratio~$R_M=\mu_pG_M^n/\mu_nG_M^p$.

\begin{figure}[!Hpb]
\begin{center} 
\includegraphics[clip=true,scale=0.47,angle=0.]{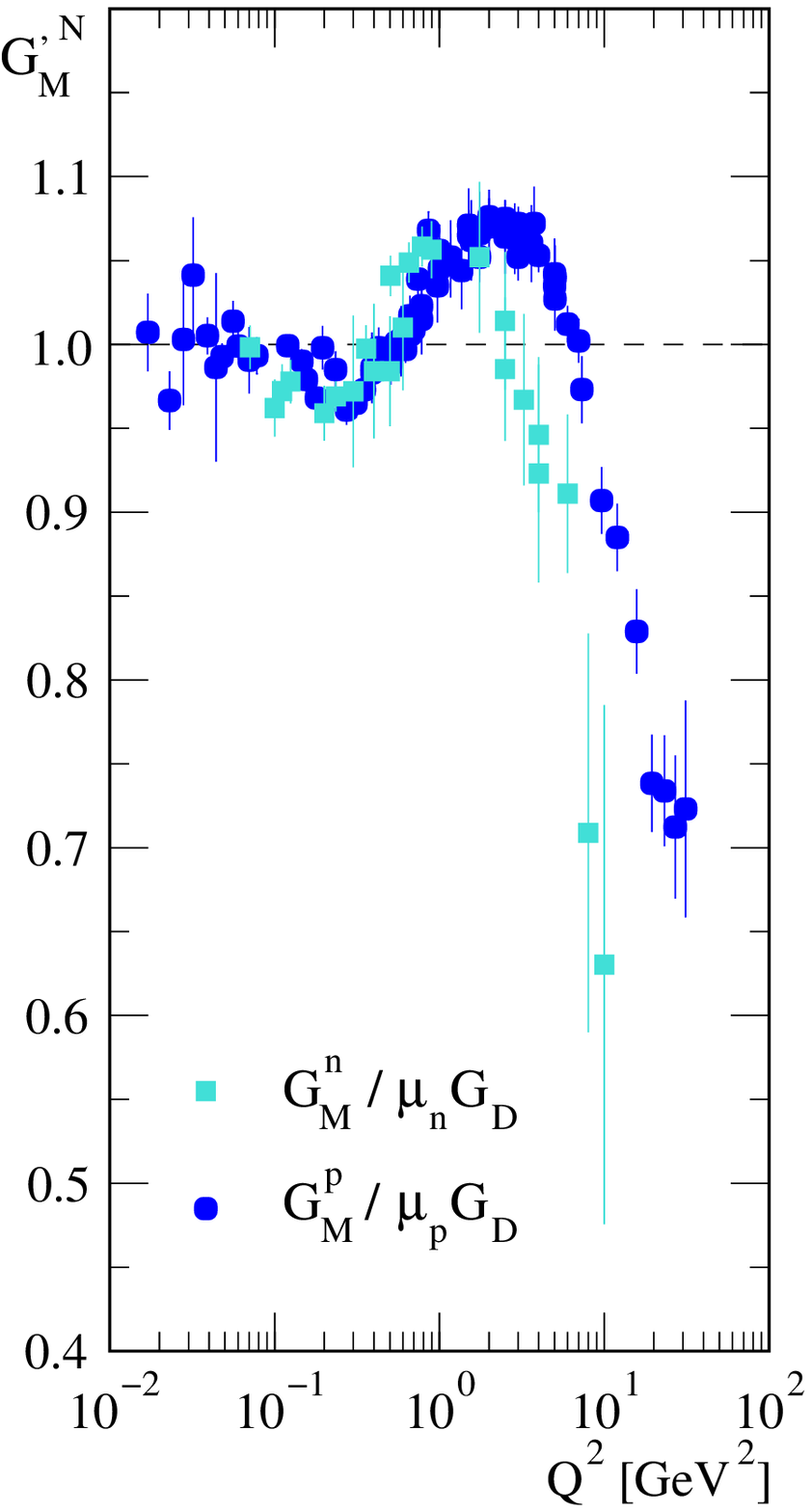}~~~~
\includegraphics[clip=true,scale=0.47,angle=0.]{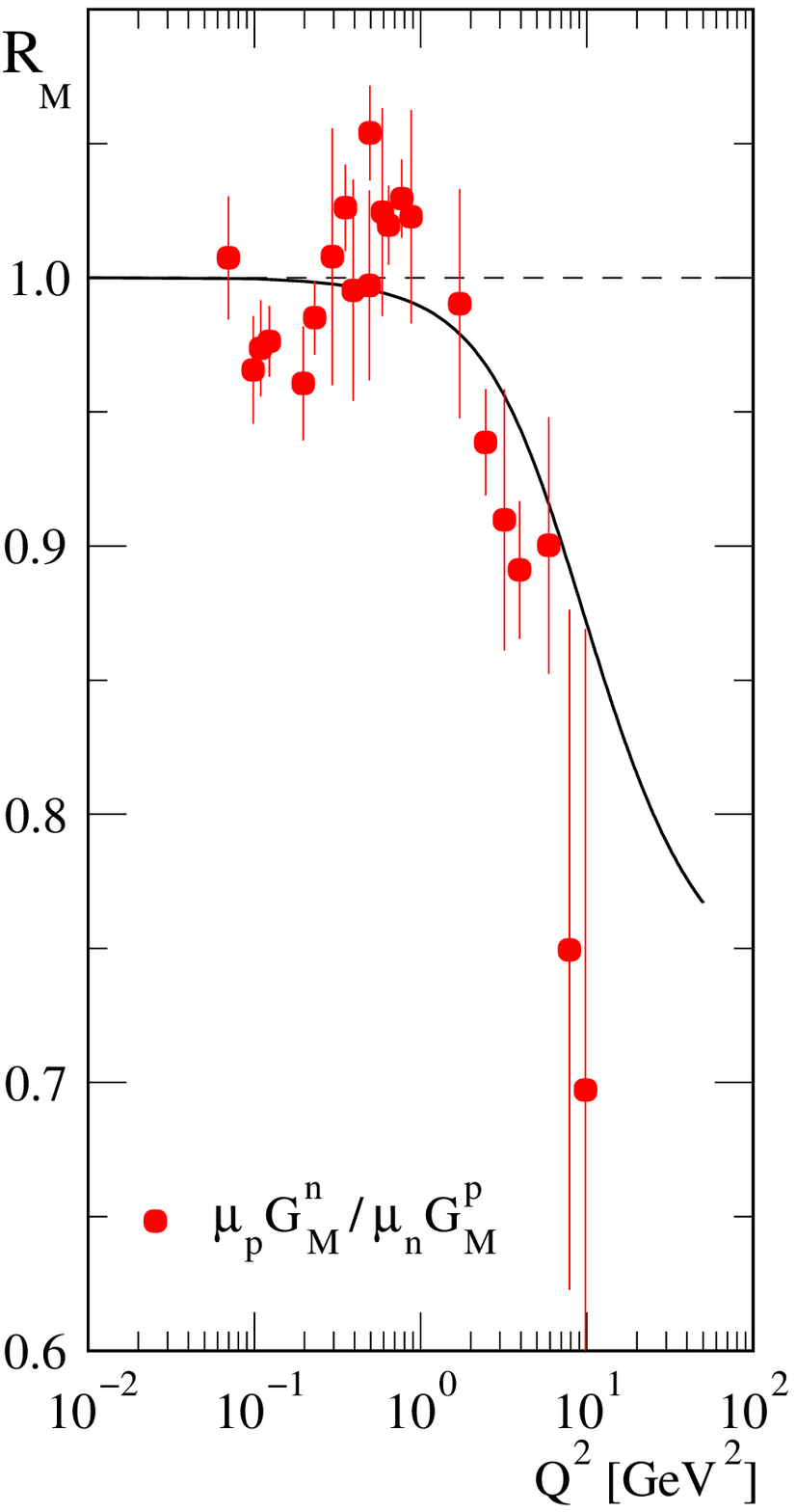}
\caption{\label{fig:ratio}
The reduced magnetic form factors~$G_M^{'N}=G_M^N/\mu{G_D}$ for proton and
	    neutron (left).
For the ratio $R_M=\mu_pG^n_M/\mu_nG^p_M$ the experimental
	    results are compared to the present
            model calculation (right).%
}%
\end{center}
\end{figure}

To this purpose we have plotted in Fig.~\ref{fig:ratio}, left, the reduced
  nucleon magnetic form factors~$G_M^{'N}=G_M^N/\mu{G_D}$, both normalized by
  their magnetic moment~$\mu$ and $G_D$ given by Eq.~(\ref{DPL}). Exactly the
  same data sets as select in Ref.~\cite{Friedrich:2003iz} are being used here
  (see Table.1 of~\cite{Friedrich:2003iz}). Within errors, the neutron
  magnetic form factor $G_M^{'n}$ follows the proton one up to
  $\sim1$~GeV$^2$, thereafter $G_M^{'n}$ decreases faster than $G_M^{'p}$.
  The left panel of Fig.~\ref{fig:ratio}, clearly shows a different shape for
  $G^{'p}_M$ and $G^{'n}_M$ in the $Q^2$ range above 1~GeV$^2$.  The different
  $Q^2$-dependences result in a deviation from unity for the ratio~$R_M$ as
  seen in Fig.~\ref{fig:ratio}, right. This ratio~$R_M$ was obtained by
  interpolating the more numerous $G_M^p$ data to the obtain the divisor
  needed at the proper $Q^2$ of the $G^n_M$ values.

In this paper we will not follow the discussion of
  Ref.~\cite{Friedrich:2003iz} concerning features below $Q^2$$\sim$1~GeV$^2$
  where pionic degrees of freedom contribute. Note, that the dip in both form
  factors at $Q^2\sim0.2$~GeV$^2$ is purely a pion-cloud phenomena.  A dip at
  low $Q^2$ was noted in the calculations of Ref.~\cite{Miller}.

As well known, the nucleon e.m. form factors are functions of the square of
 the four-momentum transfer in the scattering process:
 ${Q}^2=-q^{\mu}q_{\mu}$.  The Sachs form factors, $G_{E(M)}$, can be obtained
 from the Dirac and Pauli form factors~$\mathcal{F}_1$ and~$\mathcal{F}_2$,
 respectively, which in turn are defined through the nucleon e.m. operator
 $J^{\mu}_{em}(x)$ satisfying the requirements of relativistic covariance and
 the condition of gauge invariance. The Sachs form factors fully characterize
 the charge and current distributions inside the nucleon~\cite{Sachs1} and
 within the Breit frame the nucleon electric $G_E$ and magnetic $G_M$ form
 factors can be interpreted as Fourier transforms of the distributions of
 charge and magnetization, respectively,
\begin{eqnarray}
\Big<N_{s'}(\frac{\qvec}{2})\Big|\bm{J}_{em}(0)%
  \Big| N_s(-\frac{\qvec}{2}) \Big> &=&%
\chi^{\dagger}_{s'}\frac{i\bm{\sigma}\times\qvec}{2M_N}%
	\chi_sG_M(\qvec^2) \ \ \nonumber\\
\Big<N_{s'}(\frac{\qvec}{2})\Big|~J^{0}_{em}(0)%
  \Big| N_s(-\frac{\qvec}{2}) \Big> &=&%
   \chi^{\dagger}_{s'} \chi_s G_E(\qvec^2) \ \
\end{eqnarray}
where $\chi^{\dagger}_{s'}$ and $\chi_s$ are Pauli spinors for the initial and
  final nucleons. Note, that in the Breit frame the energy transfer vanishes
  and the incoming and outgoing momenta are $\pvec=-\qvec/2$ and
  $\pvec'=\qvec/2$; and thus $Q^2=\qvec^2$ follows.

We start our consideration of nucleon e.m. form factors from the
  non-relativistic constituent quark model~\cite{Isgur}, where the effective
  degrees of freedom are the massive quarks moving in a self-consistent
  potential whose specific form is dictated by QCD. Other degrees of freedom
  like Goldstone bosons or gluons are not considered in its original version
  and effectively absorbed into the constituent quarks. In its simplest
  version with the harmonic oscillator confining potential, the nucleon
  e.m. form factors~$G^{N}_E$ and~$G^{N}_M$ are
\begin{eqnarray} 
\label{OB_E}
G^{N}_E (\qvec^2) &=& e_N \cdot\exp\left(-\qvec^2  b^2/6 \right), \\
\label{OB_M}
G^{N}_M (\qvec^2) &=& \mu_N \cdot\exp\left(-\qvec^2 b^2/6\right)  
\end{eqnarray} 
where $e_N$ and $\mu_N$ are the charge and magnetic moment of the
 nucleon
\begin{eqnarray} 
  e_N &=& \frac{1}{2} \langle N | (1 +  \tau_3) | N \rangle, \\ 
\label{mun}
\mu_N &=& \frac{M_N}{m_q}\ \frac{1}{6} \langle N | (1 + 5\tau_3) | N \rangle 
\end{eqnarray} 
here $M_N$ and $m_q$ are the nucleon and quark mass, respectively. The
  constant~$b$ in Eqs.~(\ref{OB_E}) and~(\ref{OB_M}) determines the average
  hadronic size of the baryon and usually is called the {\it quark core
  radius}. Due to the same momentum dependence, Eqs.~(\ref{OB_E})
  and~(\ref{OB_M}) lead to the scaling law noted in Eq.~(\ref{scaling}). A
  ratio of unity is predicted by this ansatz for $R_p$ as well as for $R_M$.
  Clearly, the scaling law is in contradiction for the ratio~$R_p$,
  considering the recent experiments~\cite{protonratio}, and to $R_M$ of
  Fig.~\ref{fig:ratio}, right.

The failure of the simplest non-relativistic concept, which otherwise is quite
  successful in spectroscopy, calls for other mechanisms or other degrees of
  freedom which can generate the necessary effect. As emphasized in our recent
  study~\cite{KG}, the standard Isgur-Karl phenomenology provides such a
  possibility. The starting point is the short-ranged residual $qq$
  interaction.

The phenomenological residual interaction $V^{res}$ can be based on various
  $qq$ potentials~\cite{GlozmanRiska,Isgur}, constrained by symmetries and the
  properties of QCD; still, its dynamical origin is rather uncertain.  But
  since the effect of the residual $qq$ interaction is clearly seen in the
  excitation spectra of hadrons, one expects the corresponding interaction
  currents to play an important role for various e.m. properties of hadrons.
  In the presence of residual interactions between the quarks, current
  conservation requires that the total e.m. current operator of the hadron
  cannot simply be a sum of free quark currents, but must be supplemented by
  corresponding interaction currents. These currents are closely related to
  the $qq$ potential from which they can be derived by minimal substitution.

\begin{figure}[htb]
\begin{center}
\includegraphics[clip=true,scale=0.27,angle=0.]{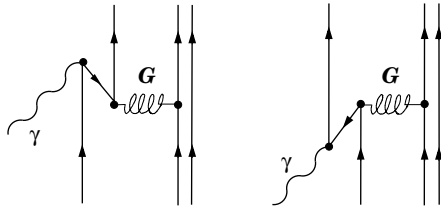}
\caption{\label{OGE}
  The OGE currents.}
\end{center}
\end{figure}

The OGE short-range potential between constituent quarks, $V^{res}=V^{OGE}$,
 can be derived from the QCD interaction Lagrangian:
\begin{equation}
\label{L_OGE}
\mathcal{L}^{QCD}(x) = - \frac{\alpha_s}{2} \bar{\psi}(x)~ \lambda_a
				\gamma_{\mu} {G}_{a}^{\mu}(x)~ \psi(x)
\end{equation}
where ${G}_{a}^{\mu}$ and $\psi$ are the gluon the quark fields, respectively,
  and $\alpha_s$ is the strong coupling constant. The explicit
  non-relativistic form of $V^{OGE}$ can be found in Ref~\cite{Thomas_book}.
  In the presence of the OGE force both, the photon and the gluons interacting
  with the quarks, can produce \qbq\ pairs leading to a pair-current
  contribution to the e.m. quark current operator. The e.m. currents we
  consider are depicted in Fig.~\ref{OGE}.
The non-relativistic reduction of these diagrams leads to the following
  OGE-induced configuration-space current operators~\cite{GrabmayrBuch} for
  charge~$\rho_{3q}^{OGE}$ and magnetization~$\jvec_{3q}^{OGE}$
\begin{eqnarray}
\label{rhooge}
\rho_{3q}^{OGE} &=& - i \frac{ \alpha_s}{16 m_q^3} \sum_{i < j}
\bm{\lambda}_i \cdot \bm{\lambda}_j \frac{ \mathcal{Q}_i}{r^3_{ij}}
\left[ e^{i \qvec\cdot \rvec_i} \Big(\qvec\cdot(\rvec_{i} - \rvec_{j})
						\right.\hfill\nonumber \\
&+&  \left.
\Big[\bm{\sigma}_i \times \qvec\Big] \Big[\bm{\sigma}_j \times
(\rvec_{i}-\rvec_j)\Big] \Big) + (i \leftrightarrow j) \right]  \hfill\\
\label{joge}
\jvec_{3q}^{OGE} &=& - \frac{\alpha_s}{8 m_q^2} \sum_{i < j}
  \bm{\lambda}_i \cdot \bm{\lambda}_j \frac{ \mathcal{Q}_i}{r^3_{ij}}
							\hfill\nonumber \\ 
&\times& 
\Big[ e^{i\qvec\cdot\rvec_i}
   \Big[(\bm{\sigma}_i + \bm{\sigma}_j ) \times (\rvec_{i}-\rvec_j)\Big]
      +(i \leftrightarrow j) \Big] \hfill
\end{eqnarray}
where~$\bm{\lambda}_i$ are the Gell-Mann $SU(3)$ colour matrices of the $i$-th
  quark normalized to {$\langle\bm{\lambda}_i\cdot\bm{\lambda}_j\rangle=-8/3$}
  for a $qq$ pair in a baryon, $\bm{\sigma}_i$ are Pauli matrices, $\rvec_i$
  is the coordinate of the $i$-th quark and $\mathcal{Q}_i$ is its charge in
  units of $e$:~ $\mathcal{Q}_i=1/2\left[1/3+\tau_3^i\right]$.  These OGE
  currents describe a \qbq\ pair creation process induced by the external
  photon with subsequent annihilation of the \qbq\ pair into a gluon, which is
  then absorbed by an another quark.  It is evident, that production of
  additional \qbq\ pairs will screen (distort) the primary distribution of the
  charge and magnetization originating due to the constituents.

As shown in Ref.~\cite{KG}, Eqs.~(\ref{rhooge}) and~(\ref{joge}) lead to the
  following electric
\begin{equation}
\label{OGE_E}
\left. \begin{array}{r} G^{OGE}_{E_p}(\qvec^2) \\
		        G^{OGE}_{E_n}(\qvec^2) \end{array} \right\} =
\ds  -\frac{\alpha_s}{m_q^3} ~q~e^{-\qvec^2 b^2 /24}
\left\{\begin{array}{r}1/3\\-2/9\end{array} \right\} \mathcal{K}(q)
\end{equation}
and magnetic contributions to the form factors
\begin{equation}
\label{OGE_M}
\left. \begin{array}{r} G^{OGE}_{M_p}(\qvec^2) \\
		        G^{OGE}_{M_n}(\qvec^2) \end{array} \right\} =
\ds   \frac{\alpha_s}{m_q^2} \frac{M_N}{q}~e^{-\qvec^2 b^2 /24}
\left\{\begin{array}{r}2/3\\-2/9\end{array} \right\} \mathcal{K}(q)
\end{equation}
where the function $\mathcal{K}(q)$ can be expressed
 analytically in terms of generalized Hyper-Geometrical functions
 $_sF_t(\alpha_1,\cdots,\alpha_s;\beta_1,\cdots,\beta_t;z)$:
\begin{eqnarray}
\mathcal{K}(q) = \frac{q}{12b}\sqrt{\frac{2}{\pi}}
\Big[ 3 \ {_2F_{2}}(1,1;\frac{3}{2},2; z) 
	- {_2F_{2}}(1,1;2,\frac{5}{2}; z) \Big] \nonumber 
\end{eqnarray}
with $z=-b^2\qvec^2/8$ and $q=|\qvec|$.

Recently~\cite{KG} we have suggested, that the dynamical $SU(6)$ breaking
  induced by gluonic currents and the resulting screening corrections can be
  considered as a possible mechanism of the scaling law violation in the
  proton ratio~$R_p$ at momentum transfers $Q^2$ beyond 1~GeV$^2$, where the
  soft pion cloud is assumed to be of lesser importance.  Using the
  prescriptions of Ref.~\cite{Licht} for the Lorentz boost of the nucleon wave
  function, Eqs.~(\ref{OB_E}), (\ref{OB_M}), (\ref{OGE_E}) and~(\ref{OGE_M})
  were taken to show~\cite{KG}, that the effect of gluonic currents to the CQM
  is important, and that the ratio~$R_p$ can be well reproduced by the
  $SU(6)\times O(3)$ wave function from the NRCQM. For the neutron ratio
  $R_n=\mu_nG_E^n/G_M^n$, this mechanism leads to the prediction, that $R_n$
  rises when the proton ratio $R_p$ falls with increasing momentum transfer
  $Q^2$ obeying our scaling relation
  $\mu_nG_E^n/G_M^n\simeq\frac{2}{3}~(1-\mu_pG_E^p/G_M^p)$~\cite{KG}. Obviously,
  due to the recent progress in using polarized nuclear targets, this
  statement can be verified experimentally.

Now we come to the another aspect of gluonic currents, namely to the scaling
  law violation in the ratio $R_M=\mu_pG_M^n/\mu_nG_M^p$.  As one can see from
  Eqs.~(\ref{OGE_E}) and~(\ref{OGE_M}), the key for the decrease of the proton
  ratio~$R_p$ relies on the opposite signs of the screening corrections to the
  electric (lead to reduction, i.e. charge screening) and magnetic (increase,
  i.e. magnetization anti-screening) form factors. Considering now $R_M$, the
  gluonic currents contribute again with opposite signs: they are positive for
  $G_M^p$ and negative for the $G_M^n$. Because the bare nucleon magnetic
  momenta, $\mu_p$ and $\mu_n$ in Eq.~(\ref{mun}), have also opposite signs
  (positive for the proton), the resulting effect is ``anti-screening'' in
  both cases.  At the same time the magnitude of the gluonic corrections are
  larger by a factor 3 for the proton than for the neutron Eq.~(\ref{OGE_M}),
  which implies that
\begin{equation}
\label{GnGp}
\mu_p G_{M}^{n}/\mu_n G_{M}^{p} < 1
\end{equation}
for $Q^2>0$. Clearly, the scaling law $G_M^n/\mu_n=G_M^p/\mu_p$ is violated.
  We use Eqs.~(\ref{OB_E}), (\ref{OB_M}), (\ref{OGE_E}) and~(\ref{OGE_M}) to
  examine the resulting effect and particularly its $Q^2$ dependence. Our
  results for the ratio~$R_M$ are shown by the full line in
  Fig.~\ref{fig:ratio}~(right), where we have used the model parameters
  obtained from the best fit to the proton ratio $R_p$ with $\alpha_s$=0.4,
  $b$=0.5 and $m_q$=400~MeV. The trend of the data is reproduced.

One can argue that the dynamical $SU(6)$ breaking induced by gluonic currents
  and the resulting e.m. screening are idealized pictures which can be
  justified only in the momentum transfer region where soft pionic effects are
  small~\cite{Miller}. We agree and our approach clearly indicates the
  presence of virtual $|qqq+\qbq~\rangle$ (soft) Fock components in the
  nucleon wave function due to the small value of $\alpha_s$.  The fitted
  value is in the range of $0.4<\alpha_s<0.6$, which however is not able to
  reproduce the $N-\Delta$ mass splitting in the case of pure OGE
  exchange~\cite{KG}. Note, that $\alpha_s$ in perturbative QCD (pQCD) should
  go to zero for large inter-quark momenta.  Here and in Ref.~\cite{KG}, we
  use $\alpha_s$ as an effective and momentum independent constant.  Following
  Ref.~\cite{Thomas_book}, we have proposed that the observed mass splitting
  can be the result of a linear combination of the pion-loop and OGE
  contributions. Consequently, the pionic contributions can produce the
  desirable effect of reducing the size of the strong coupling
  constant~$\alpha_s$.

Another important statement which can be verified, if the ratio~$R_M$ will be
  measured with much higher precision, is the role of quark configurations
  with non-zero orbital angular momentum or equivalently the possible role of
  the nucleon deformation~\cite{Miller:2003sa}.  Note, that the dynamical
  $SU(6)$ breaking induced by OGE do not contradict hadron helicity
  conservation -- this holds at least for our framework, because of the
  non-relativistic structure of nucleon wave function. At the same time, when
  imposing Poincar\'e invariance in a relativistic CQM causes substantial
  violation of the helicity conservation rule~\cite{MillerFrank}, and the
  resulting asymptotic behaviour of form factors differs from that expected in
  pQCD~\cite{Lepage}.  By now it is well established that the behavior of the
  ratio of Pauli to Dirac form factors, $QF_2(Q^2)/F_1(Q^2)$, which is
  approximately constant in this $Q^2$ range, indicates the presence of quarks
  in the proton with non-zero orbital angular momentum. We propose to consider
  the ratio~$R_M$ and its decrease as an additional test for the models where
  relativistic effects are generated by kinematical SU(6) breaking due to the
  Melosh rotation of the constituent spins~\cite{MillerFrank,Cardarelli}.

\begin{figure}[ht]
\begin{center} 
\includegraphics[clip=true,scale=0.47,angle=0.]{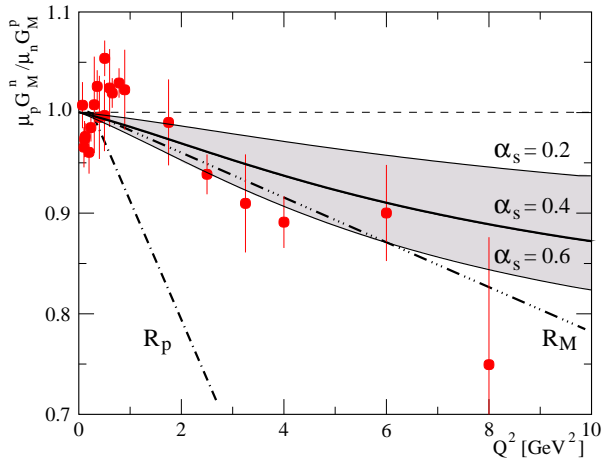}
\caption{\label{fig:RatioLinear}
The data and theoretical curves show a nearly linear decrease of the ratio
  $R_M=\mu_pG^n_M/\mu_nG^p_M$ with $Q^2$. The grey area between the two solid
  curves indicates the range of $R_M$ due to the variation of parameters
  obtained from Ref.~\cite{KG} by fitting the ratio $R_p=\mu_pG_E^p/G^P_M$.
}
\end{center}
\end{figure}
Furthermore, the nearly linear decrease of $R_M$ with increasing~$Q^2$ is
   demonstrated by the dash-triple-dotted line obtained from a fit to the data
   in Fig.~\ref{fig:RatioLinear}. This decrease is slower than that for $R_p$
   (dot-dashed line).  Finally to give a measure of sensitivity, the shaded
   area indicates the variation of the calculated ratio for values of
   $\alpha_s$ within the range of 0.2 to 0.6, which are the same around the
   optimum value of 0.4 as used in the analysis of $R_p$~\cite{KG}.  Future
   studies of contributions due to the pion cloud are expected to reduce the
   spread of parameters.

In summary, we have noted a scaling law violation in the ratio of the neutron
  to proton magnetic form factors~$R_M$, deviating from unity for
  $Q^2>$1~GeV$^2$. We have proposed that $SU(6)$ breaking induced by OGE and
  the resulting e.m. screening produces this decrease, which is less
  pronounced but of the same origin as that for the proton ratio~$R_p$ seen in
  the recent polarization transfer measurements from JLab. The good
  description by the model supports the idea of a different spatial
  distribution of magnetization inside neutron and proton. We also mention,
  that our results are instructive, both theoretically and experimentally,
  suggesting further checks of our statements against other hadronic
  models. Certainly, to confirm our results, new data on neutron and proton
  magnetic form factors with much higher precision in the momentum transfer
  region $Q^2>1$~GeV$^2$ are needed.  This will allow to employ the ratio
  $R_M$ in combination with $R_p$ as a powerful test for any theoretical
  models of nucleon structure.

\begin{acknowledgments}
Very useful discussions with G.A.~Miller are gratefully acknowledged.  This work
  was supported by the Deutsche Forschungsgemeinschaft under contracts
  Gr1084/3, He2171/3 and GRK683.
\end{acknowledgments}

\end{document}